\documentclass[aps,showpacs]{revtex4}

\begin{document}

\title{ Reply to the comments by I. Angeli and J. Csikai on : \\ `Cross sections of neutron-induced reactions' }

\author{ Tapan Mukhopadhyay$^1$, Joydev Lahiri$^2$ and D. N. Basu$^3$ }

\affiliation{Variable  Energy  Cyclotron  Centre, 1/AF Bidhan Nagar, Kolkata 700 064, India }
\email[E-mail 1: ]{tkm@veccal.ernet.in}
\email[E-mail 2: ]{joy@veccal.ernet.in}
\email[E-mail 3: ]{dnb@veccal.ernet.in}

\pacs{24.10.-i, 25.40.-h, 25.60.Dz, 28.20.Cz}

\maketitle

    The first equation of Eqs.(3) in \cite{Mu10} was used to describe the mass number and energy dependence of experimental total neutron cross sections for the first time in \cite{An70}, while the second and third ones for scattering and reaction cross sections in \cite{An74}. We are sorry for the omission of these two references which were not in our knowledge. In fact we derived these equations and Eq.(4) of Ref.[12] [J.D. Anderson and S.M. Grimes, Phys. Rev. {\bf C 41}, 2904 (1990) \cite{An90}] of our paper \cite{Mu10} as follows. From partial wave analysis of scattering theory, we know the standard expressions for scattering $\sigma_{sc}$ and reaction $\sigma_r$ cross sections as

\begin{equation}
\sigma_{sc}=\frac{\pi}{k^2} \Sigma_l~(2l+1)|1-\eta_l|^2,~~~~\sigma_r=\frac{\pi}{k^2} \Sigma_l~(2l+1)[1-|\eta_l|^2]
\label{seqn5}
\end{equation}
\noindent
where the quantity $\eta_l = e^{2i\delta_l}$. With the assumption that the phase shift $\delta_l$ is independent of $l$ and the summation over partial waves $l$ is upto $kR$ only, it follows that $\sigma_{sc}=\pi(R+\lambdabar)^2(1+\alpha^2-2\alpha \cos\beta)$, $\sigma_r=\pi(R+\lambdabar)^2(1-\alpha^2)$ and $\sigma_{tot}= \sigma_{sc} + \sigma_r = 2\pi(R+\lambdabar)^2 (1-\alpha \cos\beta)$ where $\lambdabar=1/k$, $R$ is the channel radius beyond which partial waves do not contribute, $\beta=2 {\rm Re}\delta_l = 2 {\rm Re}\delta $, $\alpha = e^{-2 {\rm Im}\delta_l} = e^{-2 {\rm Im}\delta}$ and summing over $l$ from 0 to $kR$ yielded $\Sigma_l~(2l+1) = (kR+1)^2$. \\

    We used the name `nuclear Ramsauer model' from Ref.[12] of our paper \cite{Mu10}. Carpenter \cite{Ca59} was the first to call the structure found in total neutron cross sections as nuclear Ramsauer effect. This name was adopted by subsequent authors although the nature of the oscillation in fast neutron cross sections is essentially different from that observed for slow electrons by Ramsauer. In other works the name `semiclassical optical model' \cite{An74} or `diffraction effect' \cite{La53} were used which are more appropriate. \\

    In fact the radius of the potential well is just $r_0 A^\frac{1}{3}=r_1 A^{\frac{1}{3}+\gamma}$ and $r_1$= constant. The parameter $\gamma$ is a very small number (0.00793) compared to $\frac{1}{3}$ needed for fine tuning. It should, therefore, be emphasized that, as mentioned in our paper \cite{Mu10}, it is $r_0$ which is used for fixing $\beta_0$. It is the channel radius which is energy dependent. Channel radius is the radius [appearing in Eqs.(3) of our paper] beyond which no partial waves contribute. It is well known from R-matrix theory that the channel radius is less than the nuclear (potential) radius which is precisely the case here. \\

    The drawback of Peterson's derivation is that the neutron (although massive) is treated like a photon and as its velocity inside nucleus and vacuum are proportional to $\sqrt{E+V}$ and $\sqrt{E}$, respectively, it would result in bending of the ray (as in optics) away from the normal inside nucleus (in fig.16 of Peterson's paper \cite{Pe62}, just the opposite was shown) where velocity is more. This would lead to the existence of critical angle $sin^{-1}\sqrt{E/(E+V)}$ beyond which there is no transmission (even in an attractive potential of $-V$!) and a refractive index less than vacuum for the nuclear medium which are physically unacceptable. Even then if one sticks to Peterson's assumption of a light ray, the average chord length inside nucleus, with ray bending away from the normal, turns out to be less than our result of $4R/3$ as opposed to greater than $4R/3$ as derived in Peterson's paper where $R$ is nuclear radius. However, his result \cite{Pe62} goes over to our result of $4R/3$ asymptotically at energies higher than magnitude $V$ of the real part of the nuclear potential. \\

    Obviously, these omissions do not affect the results and conclusion in the original manuscript \cite{Mu10}. \\

    We thank Drs. I. Angeli and J. Csikai for bringing this matter to our attention. 

\pagebreak

\end{document}